# Astro2020 Science White Paper

# All-Sky Near Infrared Space Astrometry

**Thematic Areas:** ☒ Planetary Systems  ☒ Star and Planet Formation  ☒ Formation and Evolution of Compact Objects  ☒ Cosmology and Fundamental Physics  ☒ Stars and Stellar Evolution,  ☒ Resolved Stellar Populations and their Environments  ☒ Galaxy Evolution  ☒ Multi-Messenger Astronomy and Astrophysics

**Principal Author:**
Name:           Barbara McArthur
Institution:    McDonald Observatory, University of Texas at Austin, Austin, TX 78712-1205
Email:          mca@astro.as.utexas.edu
Phone:          (512)471-3411

**Co-authors:** (names and institutions)
David Hobbs, Lund Observatory, Box 43, 22100, Lund Sweden
Erik Høg, Niels Bohr Institute, Blegdamsvej 17, 2100 Copenhagen Ø, Denmark
Valeri Makarov, US Naval Observatory, Washington DC, USA
Alessandro Sozzetti, INAF-Osservatorio Astrofisico di Torino, Via Osservatorio 20, 10025 Pino Torinese, Italy
Anthony Brown, Leiden Observatory, Leiden University, Niels Bohrweg 2, 2333 CA Leiden, the Netherlands
Alberto Krone Martins, CENTRA/SIM Faculty of Science, University of Lisbon, 1749-016 Lisbon, Portugal
Jennifer Lynn Bartlett, U.S. Naval Observatory, Washington, DC 20392
John Tomsick, Space Sciences Lab, UC Berkeley, Berkeley, CA 94720
Mike Shao, NASA Jet Propulsion Laboratory, Pasadena, CA 91109
Fritz Benedict, McDonald Observatory, University of Texas at Austin, Austin Tx 78712
Eduardo Bendek,  NASA Ames Research Center, Moffett Field, CA, 94035
Celine Boehm, Physics, University of Sidney, Sydney, Australia
Charlie Conroy, Harvard & Smithsonian Center for Astrophysics, Cambridge, MA 02138
Johan Peter Uldall Fynbo, Cosmic Dawn Centre, Niels Bohr Institute, Copenhagen University, Lyngbyvej 2, DK-2100 Copenhagen O, Denmark
Oleg Gnedin, Astronomy, University of Michigan, Ann Arbor, MI 4810
Lynne Hillenbrand, Astronomy, California Institute of Technology, Pasadena, CA 91125
Lennart Lindegren, Lund Observatory, Lund University, Box 43, 22100 Lund, Sweden
David R. Rodriguez, Space Telescope Science Institute, Baltimore, MD 21218
Rick White, Space Telescope Science Institute, Baltimore, MD 21218
Slava Turyshev, NASA Jet Propulsion Laboratory, Pasadena, CA 91109
Stephen Unwin, NASA Jet Propulsion Laboratory, Pasadena, CA 91109
ChengXing Zhai,  NASA Jet Propulsion Laboratory, Pasadena, CA 91109

**Abstract**

Gaia is currently revolutionizing modern astronomy. However, much of the Galactic plane, center and the spiral arm regions are obscured by interstellar extinction, rendering them inaccessible because Gaia is an optical instrument. An all-sky near infrared (NIR) space observatory operating in the optical NIR, separated in time from the original Gaia would provide microarcsecond NIR astrometry and millimag photometry to penetrate obscured regions unraveling the internal dynamics of the Galaxy. Such an observatory, GaiaNIR (Hobbs et al., 2016), was proposed which would improve Gaia proper motions errors by factors of 14–20, extend the baseline for long-period exoplanets and binaries beyond Gaia, and maintain the slowly, but continuously, degrading accuracy of the Gaia optical reference frame well beyond the mid-21st Century. These two all-sky space observatories would provide an astrometric foundation for all branches of astronomy—from the solar system and stellar systems, including exoplanet systems, to compact galaxies, quasars, neutron stars, black hole binaries and dark matter (DM) substructures.

Such a NIR space observatory is however impossible today: it requires new types of NIR Time Delay Integration (TDI) detectors to scan the entire sky and to measure global absolute parallaxes. Although developing TDI-NIR detectors is a significant challenge, the US is well-placed to advance such technology, and thus to open the doors to what can become the first international collaboration for a global astrometry space observatory as was advocated by Høg (2018) in a widely distributed message. In 2017, a European Space Agency (ESA) study of the GaiaNIR proposal already hinted that a US-European collaboration would be the optimal answer to make GaiaNIR science and technology a reality.

**Introduction**

The current ESA Gaia mission has only just begun to revolutionize our understanding of the Galaxy. In April 2018, the second data release gave 5-parameter astrometry for more than 1.3 billion sources while subsequent releases will give increasingly accurate and comprehensive sets of astrophysical data. Gaia will eventually provide positions, absolute parallaxes, and proper motions to unprecedented accuracies (20–25 µas yr$^{-1}$ at G=15) with the addition of all-sky homogeneous multi-color photometry and spectroscopy. These unique capabilities go well beyond and at the same time are highly complementary to the science cases being addressed by the multiple ground-based surveys as RAVE, SDSS, Pan-STARRS, APOGEE, LSST, etc. Gaia operates in optical wavelengths, and thus it is blind to several physical processes taking place within the obscured region of the Milky Way. A mission in NIR (at a bandpass of 400-1800 nm), however, would be able to unravel these processes, and would also multiply the number of observed objects by ~2.5 down to G=21 mag.

An all-sky absolute astrometry GaiaNIR mission would be able to probe through the Galactic dust to observe the structure and kinematics of the star forming regions in the disk, the spiral arms and the bulge region and thus derive model independent distances and proper motions in these obscured parts of the sky. As a bonus, such new mission launched with an interval of ~20 years from Gaia, would allow a new joint solution with the old Gaia observations, which combined would originate in improved proper motions with 14–20 times smaller errors, enabling thus new kinematical and dynamical inferences about the underlying physics of the Milky Way and of objects as stellar clusters and dwarf galaxies. Parallaxes would also be improved in such joint solutions by a factor of √2 assuming two missions of equal duration. The accuracy of the new mission should be at least that of Gaia using tried and trusted instrumentation, techniques of global absolute astrometry, and some lessons learned from Gaia, e.g. improved basic angle monitoring. To achieve these goals, we need to make progress in the development of NIR/optical detectors.

**NIR astrometry and photometry**

What is the dynamics at the innermost regions of the Milky Way, deep inside the galactic plane, and at the bones of its spiral arms? The galactic center, the plane and the spiral arm regions are heavily obscured by interstellar extinction so a NIR mission is needed to map these regions astrometrically for the first time and derive their kinematics and dynamics. Linking the individual motions of obscured objects with those from the Gaia optical survey would give a highly complete picture of the dynamics of the Galactic interior at large scales, something that is impossible to attain with present interferometry-based instruments that can only survey small fields and a small number of objects. The complex nature of the Galaxy with its bulge, bar, and spiral structure can excite stars to migrate radially and induce disk heating (Friedli, Benz, & Kennicutt, 1994; Sellwood & Binney 2002), thus accurate measurements of the three-dimensional motion, distances and properties of a significant numbers of these obscured stars at the widest possible scales in the Milky Way are needed to trace the dynamical history and evolution of our Galaxy. Classical tracers as stellar clusters can be detected and studied in NIR deeply inside the disk, and can be used to probe the Galactic disk structure, kinematics and stellar formation rate and properties (e.g. Kuhn et al. 2015, 2019). But a major backbone of the Milky Way history is dark matter, and to study the dark matter properties in the inner disk direct NIR measurements are necessary. These measurements can reveal for instance whether the Galaxy has a cored or cusped dark matter halo (Governato et al. 2012; Theia Collaboration, 2017); whether there are thin, disk-like components of dark matter; and, whether the spiral arms have their own dark matter components. This would help to understand our Milky Way, and also the very nature of the elusive dark component of contemporary physics that, for the moment, only reveal itself through its impact on baryonic matter kinematics and dynamics.

A NIR-capable global astrometric mission would be able to detect very faint red objects in extinction regions, including Red Dwarfs (RD), cool White Dwarfs (WD), Brown Dwarfs (BD) and free-floating planets, which is impossible to Gaia. It would detect such objects within a large volume of the Galaxy, and also in the crucial extinction regions, such as very young open clusters and star-forming regions. These data would enable understanding of the Initial Mass Function (IMF) and the gravitational processes and conditions triggering star formation. At the last stages of stellar evolution, the study of WDs provides key information about astronomy and physics, and a NIR facility would improve the detection and characterization of the cool WD population compared to the limited reach of Gaia due to its optical operating wavelength (Carrasco et al. 2014). The comparison between empirical and theoretical luminosity functions of WDs can lead to the derivation of the Galactic age and the reconstruction of the history of the Galaxy star formation rate.

High-precision NIR astrometry will have an impact on a wide variety of stellar physics topics. For instance, additional detections of free-floating planets could reveal whether they are ejected from planetary systems or formed in collapsing dust clouds like stars. Global NIR micro-arcsecond astrometry will particularly reveal important dynamical information in the highly obscured regions that are the birthplaces of most stellar clusters and associations. Although Gaia is extending our current knowledge of clusters, most within its reach are either relatively nearby or are located at high galactic latitudes that may not be fully representative of the conditions in the innermost regions of the galactic disk. The binary orbits of exotic objects, such as neutron stars and stellar-mass black holes, will also benefit from improved mass estimates that may allow their equations-of-state to be accurately constrained for the first time (Özel & Psaltis 2009).

GaiaNIR will also enable extensive local tests of stellar standard candles and their period-luminosity relations. The key uncertainties of variable extinction and metallicity are significantly reduced at longer wavelengths, and Asymptotic Giant Branch (AGB) stars, including Mira variables and OH/IR stars, are very bright in the NIR. Mira stars are much more numerous than Cepheids. However, the period-luminosity relation of OH/IR stars is not well known. The better characterization of these standard candles through absolute and global NIR astrometry will improve the distance scale that set one of the rules for the scale of our Universe.

**Improved proper motions and parallaxes**

Gaia is a remarkable mission for detecting and characterizing nearby streams that cross the disk of the Milky Way, but it will not be sufficient to discover most of the stream-like structures in the halo. A future mission combined with Gaia could derive accurate proper motions, a factor of 14–20 better than Gaia alone, and the improved parallaxes needed to reach such larger distances. Dynamical studies in the outer halo would be greatly enhanced, resolving tangential motions in streams and local dwarf galaxies, with a potential accuracy of 2–3 km s$^{-1}$ for samples out to ~100 kpc. This will provide great insight into the gravitational potential of the outer Milky Way and its halo. Finding gaps in the streams may reveal the influence of dark matter sub-haloes lying in the halo as predicted by the ruling dark matter scenario today, and allow us to determine the dark matter distribution at large radii, the flattening of the potential, and the total mass of the Galaxy.

Hyper-Velocity Stars (HVSs) originate from gravitational interactions with massive black holes. Very accurate proper motion measurements are a key tool for studying these objects. When combined with radial velocities, the three-dimensional space velocity is obtained. Unfortunately, known HVSs are distant and on largely radial trajectories. Some HVSs originate in the Galactic center while others have an origin in the disk; however, others may have origins in the Magellanic Clouds or beyond. Accurate proper motions are needed to reconstruct their trajectories and distinguish between the different possible origins. Gnedin et al. (2005) showed that precise proper motions of HVSs would constrain the structure (axis ratios and orientation of triaxial models) of the Galactic halo. Probing more deeply into the Galactic center with NIR would detect small populations of HVSs closer to their ejection location.

Astrometrically resolving internal dynamics of nearby galaxies, as M31, dwarf spheroid galaxies, globular clusters and the Magellanic Clouds would enable a precise mapping of dark matter, including sub-structures, throughout the local group. For example, the Large Magellanic Cloud has a parallax of 20 μas, the necessary astrometric accuracy of ~10% is just within the reach of Gaia. Combining Gaia with GaiaNIR will enable accurate measurements of the internal motions of nearby galaxies and thus to astrometrically resolve dynamics within the Local Group.

**Long-period exoplanets and binary stars**

GaiaNIR can potentially explore uncharted territory in the realm of planetary-mass companions orbiting stellar and substellar primaries that cannot be observed with sufficient sensitivity using other techniques, but that are bright in the NIR. These include ultra-cool nearby dwarfs, around which GaiaNIR could complete the census of cold terrestrial planets, and derive a statistical sample of heavily reddened young stars in the cradles of the nearest star-forming regions. Systematic GaiaNIR observations of all stellar samples for planetary companions to which Gaia (particularly a fully extended mission) was originally sensitive would also bring new insight in the global architectures of planetary systems. This would complement the combined multi-technique efforts currently ongoing or planned, from the ground and in space. Astrometric binaries with components of unequal brightness can be detected for orbital periods up to ~100 years via apparent acceleration (Kaplan & Makarov 2003). Within the solar system, accurate astrometry over a longer time span would also improve our understanding of asteroid families and the mechanisms that lead to the injection of asteroids on Earth-crossing trajectories.

**The celestial reference frame**

The Gaia optical reference frame based on quasars will very slowly degrade over time due to errors in its orientation and spin and due to small proper motion patterns. Additionally, the catalog accuracy will decay more rapidly due to errors in the measured proper motions. Dense and accurate reference frames are needed for the forthcoming Extreme, Giant, and Overwhelming telescopes, also for smaller instruments and also for relative-to-absolute transformations. The detection of the reference frame quasars solely from zero linear proper motion and parallax is possible (Heintz et al. 2018), reducing the need for spectra. At microarcseconds, Quasars will have some apparent proper motions due to time-dependent source structure, and these non-linear motions can be used to either study these interesting objects and to select the most well-behaved to the construction of the reference frame.

An important aspect of a reference frame is linking it through absolute coordinates to reference frames at other wavelengths, producing reference grids that can be accessed by various surveys. This science objective lies at the heart of fundamental astrometry, and it requires maintaining the accuracy of the reference frame through time. The Gaia celestial reference frame will immediately supersede any other optical reference frames and will be the standard optical reference frame for the astronomical community for many years. Nevertheless, a new global, Gaia-like mission will be necessary to maintain this realization of the celestial reference frame (CRF). By moving towards NIR astrometry the frame will become dense in obscured regions and will provide a link to the International Celestial Reference Frame (ICRF) at new wavelengths.

Beyond Astronomy, practical applications of the fundamental CRF are mostly related to accurate positioning, navigation, and space awareness. The CRF provides the basis for the Earth orientation and Global Positioning System (GPS) implementations. These systems that operate in the optical/IR and radio domains, place increasingly demanding requirements on the accuracy of the link between the quasar-based International ICRF3 and the Gaia Reference Frame. Currently, only a few more than 2000 radio-loud quasars observed with Gaia can be used to link the two. A NIR Gaia can greatly increase the number of suitable CRF objects, especially at redshifts greater than 4, where the lack of bright Lyman series lines in the optical band significantly redden the sources. The link will also be much improved in the zone of avoidance along the Galactic plane, where few bright quasars are known due to extinction and crowding.

Even quasars, the most distant celestial objects, have nonzero proper motion. The expected systematic dipole pattern of all celestial objects' apparent motion is caused by the acceleration of the Solar System in the Galactic gravitational field and the related secular aberration (Kopeikin &

Makarov 2006). Because the amplitude of this proper motion pattern is small (4-5 μas yr$^{-1}$), Gaia is unlikely to measure it adequately. However, GaiaNIR will measured it accurately, because the secular aberration pattern will manifest itself as a >80 μas distortion of celestial positions over the 20-year time span. Other hypothetical components of nonlinear observer's motion can be experimentally tested, as the acceleration of the Galaxy with respect to the Cosmic Microwave Background and the presence of a gravitational attractor in the vicinity of the Sun.

**The need for NIR detectors**

To achieve these goals, however, we need to explore the feasibility and technological developments needed to manufacture space-qualified optical/NIR (400–1800 nm) passively cooled detectors that do not yet exist. The fundamental principle of Gaia-like observations is that the spacecraft must have a constant rotation resulting in a moving image, which must be compensated for. After the ESA GaiaNIR study there is a consensus that the TDI NIR approach is necessary for such a mission to be successful. Developing TDI in NIR detectors may be a difficult challenge, but the US has a large heritage in NIR detector technology. The quality of photo-detectors with combined optical/NIR sensitivities has improved remarkably over the past decade, including their ability to withstand radiation damage. The current leading NIR detectors in the field are HgCdTe, also known as MCT, but are unable to transfer charge like CCDs which is needed for TDI mode. One possible hybrid solution is to connect a detection layer of HgCdTe to a standard silicon-type CCD substrate chip that is used primarily for charge storage and transfer, as opposed to charge generation. Although this technology seems plausible in principle, the science drivers before a possible Gaia-like mission operating in NIR have not been sufficient to justify a detailed investigation. Other options for achieving low-noise TDI in the optical/NIR are HgCdTe CMOS devices with improved charge transfer technology or Silicon-Germanium based CCD-like devices which reach into the NIR. Both these options show promise but also require significant technology development and laboratory experiments to determine their properties and potential for GaiaNIR. With this in mind we are currently preparing a Notice Of Intent (NOI) to submit white papers on activities, projects, or state of the profession considerations to the Astro2020 Decadal Survey specifically related to investigating the detector technology.

**Outlook**

The Euclid and the LISA mission demonstrates how a NASA and ESA partnership can make new science missions successful, and the GaiaNIR mission could follow this collaboration, enabling together the future of Space Astrometry. The science cases for a NIR successor to Gaia are highly compelling. While the launch date for such mission may seem distant, work must begin now to ensure that the essential technology is ready. ESA already studied the principles of global astrometry and the technical implementation of the mission and found that GaiaNIR is feasible provided suitable focal plane detector technology is developed. As the undisputed world leaders in NIR detector technology, the US should lead the detector development. Collaborating with European colleagues on a global astrometry mission would engage US astronomers in world-class astrometry while helping to answer a wide variety of contemporary astronomical questions and strengthening the very foundations of 21st century astronomy.